\documentclass{ws-procs975x65}

\begin{document}

\title{PINNING AND BINDING ENERGIES FOR VORTICES IN NEUTRON STARS: COMMENTS ON RECENT RESULTS}

\author{P. M. PIZZOCHERO$^*$ }

\address{Dipartimento di Fisica, Universit\`a degli Studi di Milano,
  \\ and Istituto Nazionale di Fisica Nucleare, sezione di Milano,\\
  Via Celoria 16, 20133 Milano, Italy\\
$^*$E-mail: pierre.pizzochero@mi.infn.it}

\begin{abstract}
We investigate when the energy that pins a superfluid vortex to
the lattice of nuclei in the inner crust of neutron stars can be
approximated by the energy that binds the vortex to a single
nucleus. Indeed, although the pinning energy is the quantity
relevant to the theory of pulsar glitches, so far full quantum
calculations have been possible only for the binding energy.
Physically, the presence of nearby nuclei  can be neglected if the
lattice is  dilute, namely with nuclei sufficiently distant from
each other. We find that the dilute limit is reached only for
quite large Wigner-Seitz cells, with radii $R_{_{\rm WS}} \gtrsim
55 $ fm; these are found only in the outermost low-density regions
of the inner crust. We conclude that  present quantum calculations
do not correspond to the pinning energies in almost the entire
inner crust and thus their results are not predictive for the
theory of glitches.
\end{abstract}


\bodymatter
\section{Pinning and binding energies}

Pulsar glitches are sudden spin-ups in the otherwise steadily
decreasing frequency of rotating magnetized neutron stars.
According to the vortex-model, glitches may represent direct
evidence for the existence of a macroscopic superfluid inside such
stars \cite{AI}. This scenario involves the inner crust of the
star, namely the density range $\rho_d \leq \rho \leq 0.6 \rho_o
$, where $\rho_d = 4 \times 10^{11}$ g/cm$^{3}$ is the neutron
drip density and $\rho_o = 2.8 \times 10^{14}$ g/cm$^{3}$ is the
nuclear saturation density. Here, the quantized vortices that form
in the superfluid of unbound neutrons to carry its angular
momentum can attach themselves to the lattice of neutron-rich
nuclei co-existing with the neutrons in the inner crust, thus
freezing part of the superfluid angular momentum. As a consequence
of the star's spin-down, hydrodynamical forces then develop
(Magnus force) which tend to detach the vortices from the lattice
in order to let the angular momentum free to decrease. If a large
number of vortices can be unpinned simultaneously and deliver
their angular momentum to the star surface, its corresponding
sudden spin-up is observed as a glitch. The process naturally
repeats itself so that part of the superfluid angular momentum is
released in discrete time steps, thus explaining the approximate
periodicity of glitches.

A crucial  microscopic  input of the model is the structure of the
nuclear lattice  at the different densities found in the inner
crust. Calculations predict a bcc structure for the coulomb
 lattice, namely each nucleus is at the center of a cubic cell of
side $\ell_p = 2 R_{_{\rm WS}}$ with nuclei at each vertex. The
bcc lattice can also be seen as a superposition of layers of cubic
cells; each layer, of thickness $\ell_p$, is  delimited by two
planes of nuclei (those at the vertices of the cells) and contains
in its middle a third plane of nuclei  (those at the center of the
cells). In Table~1 we list the results from the classical paper by
Negele and Vautherin \cite{NV} for five zones along the inner
crust. The parameters which characterize  the lattice structure
are the baryon mass density ($\rho_{_{\rm B}}$), the number
density ($n_{_{\rm G}}$) and corresponding Fermi momentum
($k_{_{\rm F,G}}$) of the gas of unbound neutrons, the radius of
the Wigner-Seitz (WS) cell ($R_{_{\rm WS}}$) and the radius of the
nucleus ($R_{_{\rm N}}$) at the center of the cell.
\begin{table} \tbl{Properties of five zones along the inner
crust, as obtained in Ref.~\refcite{NV}.}
{\begin{tabular}{@{}cccccc@{}} \toprule Zone & $\rho_{_{\rm B}}$
& $n_{_{\rm G}}$ & $k_{_{\rm F,G}}$ & $R_{_{\rm WS}}$ & $R_{_{\rm
N}}$ \\ & (g/cm$^{3}$) & (fm$^{-3}$) & (fm$^{-1}$) & (fm) & (fm)
\\
\colrule 1 & $1.5\times 10^{12}$ & $4.8\times 10^{-4}$
         & $0.24$ & $44.0$  & $6.0$  \\
2& $9.6\times 10^{12}$ & $4.7\times 10^{-3}$
         & $0.52$ & $35.5$  & $6.7$  \\
3& $3.4\times 10^{13}$ & $1.8\times 10^{-2}$
         & $0.81$ & $27.0$  & $7.3$  \\
4 & $7.8\times 10^{13}$ & $4.4\times 10^{-2}$
         & $1.09$ & $19.4$  & $6.7$  \\
5 & $1.3\times 10^{14}$ & $7.4\times 10^{-2}$
         & $1.30$ & $13.8$  & $5.2$  \\
                   \botrule
\end{tabular}}
\end{table}

The other crucial microscopic input is the vortex-lattice
interaction potential as a function of the vortex position in the
lattice, from which the pinning force could be derived. Due to the
complicated spatial geometry of the general problem, however, a
simpler approach has been followed so far which calculates only
the \emph{pinning energy}. This is defined as the difference in
energy between two vortex-lattice configurations which are
relevant to the pinning mechanism and possess a reasonable degree
of symmetry. More precisely, one starts by taking the vortex axis
 oriented along a symmetry direction of the lattice and then considering only the
interaction of the vortex with the nuclei lying in a plane
perpendicular to the vortex axis (the middle plane of a layer).
This allows to determine the pinning energy 'per site' (i.e. per
 layer). The pinning energy 'per unit length', which is the one
 relevant to the vortex-model for glitches, can then be obtained
 if one knows  the number of pinning sites (layers) per unit
 length. Statistical estimates of this quantity allow to evaluate
 pinning also for random vortex-lattice mutual directions.

 In Figure~1~(left) the black disks represent the nuclei in the
given plane, each positioned at the center of a cube. The nuclei
at the vertices of the cubes are shown with white disks; they lie
in the planes that delimit the layer, at a distance $R_{_{\rm
WS}}$ above and below the original plane. The vortex core position
is schematically represented by the dashed line and the vortex
axis, perpendicular to the plane, is shown by a star.
\def\figsubcap#1{\par\noindent\centering\footnotesize(#1)}
\begin{figure}[t]
\begin{center}
  \parbox{2.4in}{\epsfig{figure=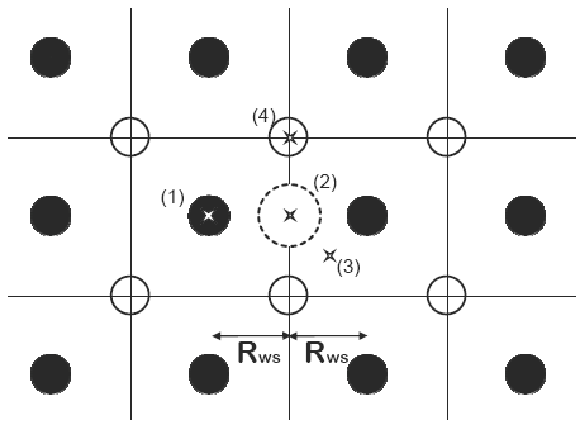,width=2.4in}}
  \hspace*{6pt}
  \parbox{2.4in}{\epsfig{figure=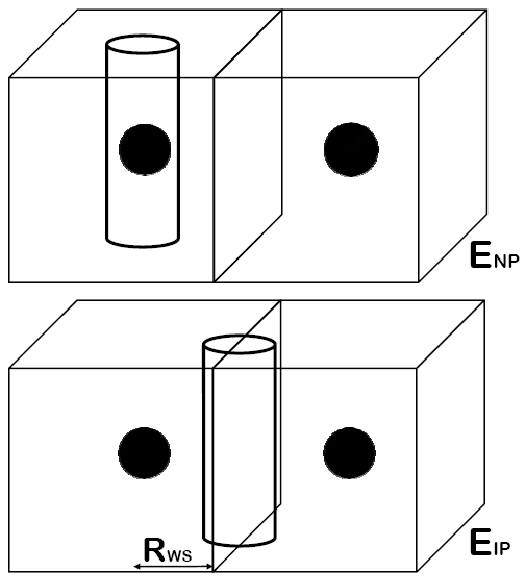,width=2.4in}}
  \caption{ Left: vortex positions in a bcc lattice.  \ Right: configurations for the
  pinning energy. }
\end{center}
\end{figure}
The nuclear pinning (NP) configuration corresponds to the vortex
pinned on a nucleus, e.g. in position (1).  The interstitial
pinning (IP) configuration corresponds to the vortex away from and
equidistant between two nearby nuclei in the plane, e.g. in
position (2). Notice that in the NP configuration the vortex also
pins to the nuclei in the layers above and below, which are
aligned along its axis with spacing $2 R_{_{\rm WS}}$.
Conversely, in the IP configuration the vortex does not pin to any
nucleus in the lattice. Position (4) is not allowed as an
interstitial pinning site; actually, it  is physically equivalent
to nuclear pinning, since the vortex will pin to the nuclei in the
layers above and below; in other words, positions (1) and (4) are
equivalent once we consider the pinning energy per unit length.

The pinning energy is the energy necessary to move the vortex by a
distance $ R_{_{\rm WS}}$, from the NP to the IP configuration.
The \emph{pinning energy per site} is then defined as $E_{\rm pin}
= E_{_{\rm NP}}-E_{_{\rm IP}}$, i.e. the difference in energy
between the two configurations. Physically, the vortex interacts
significantly only with first neighbors; therefore, it is enough
to restrict the attention to a system of two adjacent cells, as
shown in Figure~1~(right), since contributions from cells further
away will cancel out. If $E_{\rm pin} < 0$ the vortex is attracted
by the nucleus ('nuclear pinning') and the average pinning force
per site is obtained as $F_{\rm pin} = |E_{\rm pin}|/R_{_{\rm
WS}}$. If $E_{\rm pin} > 0$ the vortex is repelled by the nucleus
('interstitial pinning'); is such a case the pinning force is
several orders of magnitude smaller than $F_{\rm pin}$ and
practically negligible. This because the vortex can be moved
around the lattice completely avoiding the nuclei, e.g. from (2)
to (3), which cost almost no energy \cite{LE}.

If one considers a \emph{single} nucleus instead of the actual
lattice of nuclei, a simpler quantity can be introduced: the
\emph{binding energy} of the vortex-nucleus system is the energy
necessary to detach the vortex from the nucleus and take it to
infinity, where the background matter is uniform (unaffected by
the presence of the nucleus or the vortex). It can be defined as
the difference in energy between two configurations, one with the
vortex bound to the nucleus (i.e. with vortex-nucleus separation
$d=0$) and the other with the vortex far away and unbound from the
nucleus (i.e. with vortex-nucleus separation $d \rightarrow
\infty$). The short range nature of nuclear forces and the rapidly
decreasing kinetic potential ($\propto 1/r^{2} $) associated to
the vortex flow, however, imply that the vortex and the nucleus
interact significantly only when they are relatively close to each
other. Therefore, to a good approximation the binding energy is
reached not for $d \rightarrow \infty$, but already for some
finite separation $d = D_{\rm conv}$. The convergence distance
$D_{\rm conv}$ corresponds to a configuration presenting a zone
between the nucleus and the vortex where neutron matter has
practically reached uniformity; taking $d > D_{\rm conv}$ only
adds more uniform matter in between, whose contributions cancel
out in the binding energy.

In Figure~2~(left) the vortex and the nucleus are shown at a
distance $d$ apart. Such a system can be described by two
cylinders of radius $R = \frac{1}{2} d$ and height $h$, which are
also natural quantization boxes to evaluate the energy of this
geometry. Indeed, the vortex-nucleus bound (B) and unbound (U)
configurations are those represented in Figure~2~(right) and the
binding energy is defined as $E_{\rm bind} = E_{_{\rm B}}-E_{_{\rm
U}}$. Of course, for this to make sense the result must converge;
as explained, this will happen only when the separation is $ d
\geq D_{\rm conv}$ and thence when the cylinders have radius
larger than  $R_{\rm conv} = \frac{1}{2} D_{\rm conv}$. The
physical interpretation  of the convergence radius is the same as
before: when $R = R_{\rm conv}$, matter has practically reached
uniformity at the surface of both cylinders of Figure~2. Taking
larger boxes would only add more uniform matter at their
boundaries, whose energy contributions cancel out in the binding
energy. In this respect, we notice that the choice of cylindrical
boxes, although natural, is by no means unique; for example one
could as well take two cubic boxes of side $\ell = d$. As long as
$\ell \geq 2 R_{\rm conv}$, energy contributions from the
additional uniform matter will again cancel out in the final
binding energy.

\begin{figure}[b]
\begin{center}
  \parbox{2.4in}{\epsfig{figure=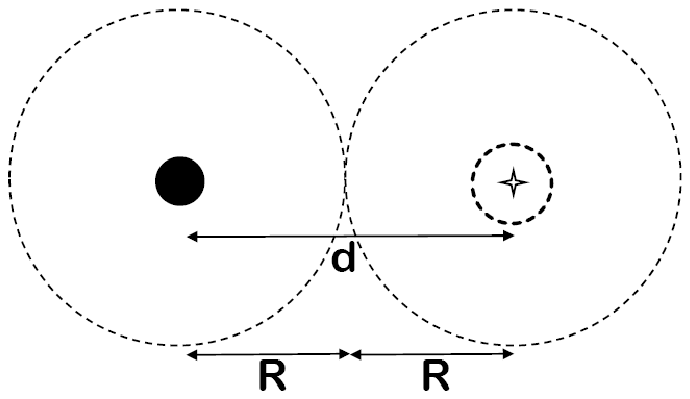,width=2.4in}}
  \hspace*{6pt}
  \parbox{2.4in}{\epsfig{figure=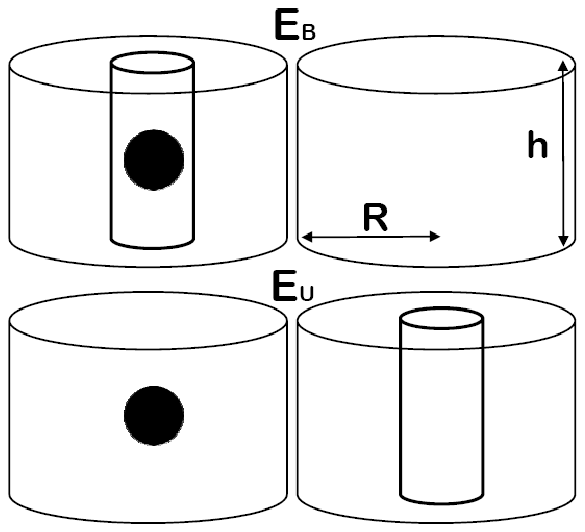,width=2.4in}}
  \caption{Left: vortex-nucleus system with separation $d$ and cylindrical quantization
  boxes of radius $R = \frac{1}{2} d$. \ Right: configurations for the binding energy. }
\end{center}
\end{figure}

The complicated geometry of the pinning configurations, with its
interplay of different symmetries (spherical for the nucleus,
axial for the vortex, periodic for the lattice), makes the
calculation of pinning energies a very difficult problem at the
quantum level. Conversely, binding energies are simpler to
calculate since the convergence requirement allows to split both
the U and B configurations into  independent axially symmetric
problems (the pairs of cylinders). The question is under which
conditions pinning and binding energies can be expected to be
equivalent.

\section{Dilute limit for the lattice}

In general, pinning and binding energies are different and
unrelated quantities. The binding energy represents the
interaction between a vortex and a single nucleus while pinning
involves the interaction of the vortex with a lattice of nuclei.
Moreover, binding is defined as the value to which the difference
$E_{_{\rm B}}-E_{_{\rm U}}$ converges for large enough
vortex-nucleus separation and thus it does not depend on any
distance parameter; conversely, pinning is defined with respect to
a specific configuration, namely the difference $E_{_{\rm
NP}}-E_{_{\rm IP}}$ must be evaluated at vortex-nucleus separation
$R_{_{\rm WS}}$, and thus it  depends crucially on the length
parameter of the lattice. Figure~3~(left) illustrates the
difference between $E_{\rm bind}$ and $E_{\rm pin}$: binding is
related only to the value of the interaction potential at the
center of the WS cells, while pinning is  determined by the local
extremum in the interaction potential which, by symmetry reasons,
must develop at the boundary between two WS cells. Whether this is
a maximum or a minimum will depend on the particular radial
dependence of the interaction potential; as long as this
dependence is not known, as it is the case for the problem under
study,  no prediction on the value or \emph{even the sign} of the
pinning energy can be extracted from the knowledge of the binding
energy, as obvious from the figure. We point out that the NP
configuration of pinning is physically equivalent to the B
configuration of binding, at least as long as the vortex core is
smaller than the WS radius so that the presence of nearby nuclei
can be neglected. Therefore, the difference between $E_{\rm pin}$
and $E_{\rm bind}$ follows essentially from the physical
difference between the IP and U configurations.

\begin{figure}[b]
\begin{center}
  \parbox{2.4in}{\epsfig{figure=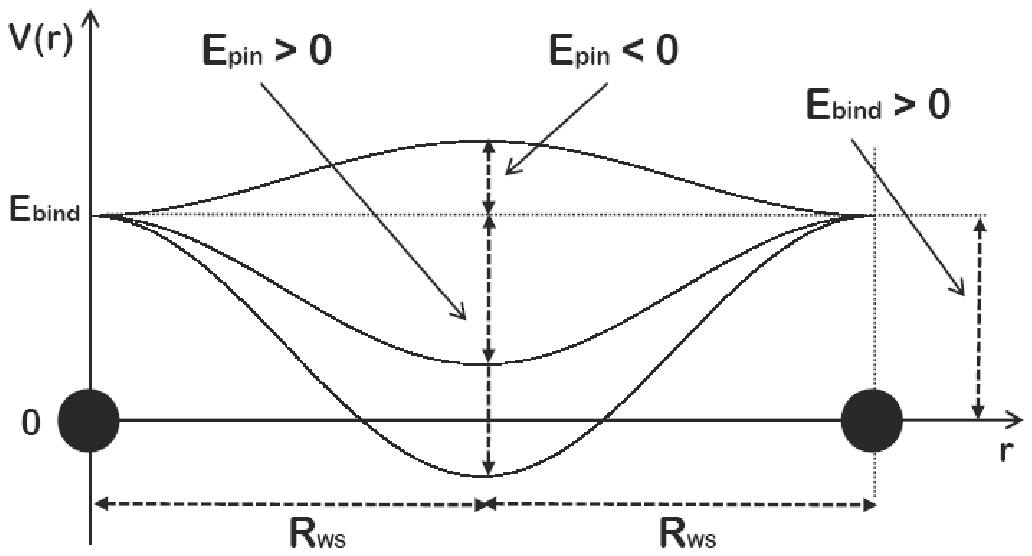,width=2.4in}}
  \hspace*{6pt}
  \parbox{2.4in}{\epsfig{figure=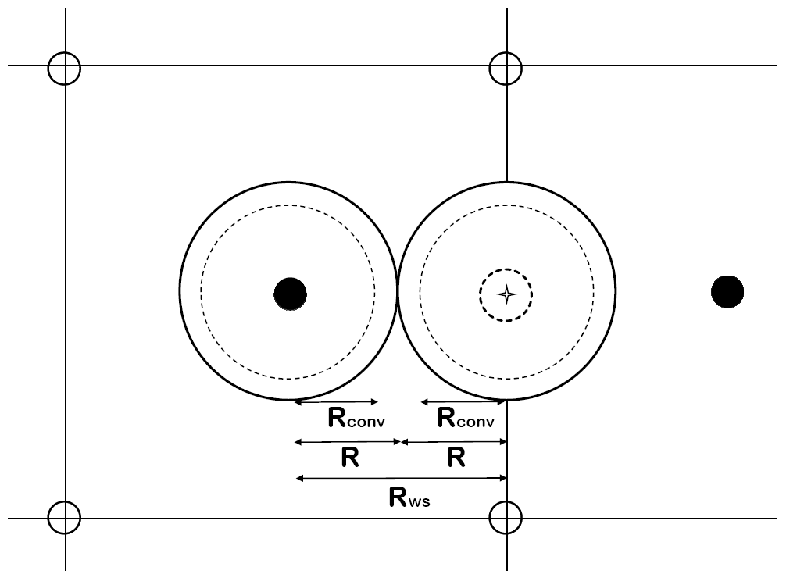,width=2.4in}}
  \caption{Left: example of different possible pinning energies
   corresponding to the same binding energy.  \  Right: lattice in the dilute limit $R_{_{\rm WS}} > 2
R_{_{\rm conv}} $. }
\end{center}
\end{figure}
If the spacing of the nuclei in the lattice is sufficiently large,
however, one expects that in the IP configuration the vortex is
sufficiently distant from the nuclei that it does not interact
significantly with them anymore, so that the IP configuration
becomes physically equivalent to the U configuration. If this is
the case, the pinning energy will coincide with the binding
energy; we call this scenario, where $E_{\rm pin} = E_{\rm bind}$,
the \emph{dilute limit} for the lattice. Quantitatively, such a
limit is reached when the vortex-nucleus distance in the IP
configuration ($R_{_{\rm WS}}$) is larger than the separation for
which they do not appreciably interact anymore ($D_{_{\rm
conv}}$).

Figure~3~(right) shows the lattice in the dilute limit, namely
with $R_{_{\rm WS}}
> D_{_{\rm conv}} = 2 R_{_{\rm conv}} $. The IP configuration with vortex-nucleus
separation $d=R_{_{\rm WS}}$ can be represented by two cylinders
(solid circles) of radius $R = \frac{1}{2} d = \frac{1}{2}
R_{_{\rm WS}} >  R_{_{\rm conv}}$. Since this radius is larger
than the convergence radius (dotted circles), the IP configuration
in this limit is indeed equivalent to the U configuration (cf.
Figure~2) and the pinning energy thus calculated is the same as
the binding energy. Conversely, if $R_{_{\rm WS}} <  D_{_{\rm
conv}} = 2 R_{_{\rm conv}} $ the lattice is in the dense limit and
$E_{\rm pin} \neq E_{\rm bind}$. Figure~4~(left) illustrates this
scenario; the cylinders needed to represent the IP configuration
have radii smaller than $R_{_{\rm conv}}$ and therefore this
geometry is not equivalent to the U configuration, which requires
convergence to be properly defined. Consequently, the energy
difference $E_{\rm pin}$ calculated from this geometry cannot be
equal to $E_{\rm bind}$, the latter  being reached only for larger
radii of the cylinders.

In conclusion, the important parameter to be determined is the
convergence distance $ D_{_{\rm conv}}$ for binding. A lower limit
can easily be found by considering the kinetic energy contribution
of a nucleus added to the vortex flow at distance $d$ from the
vortex axis \cite{EB}. This energy becomes negligible only for
vortex-nucleus separations larger than $d \approx 30$ fm
\cite{DP1}, which implies that $ D_{_{\rm conv}}$ cannot be
smaller than this value.

\section{Comments on recent results}

To date, the only consistent and realistic calculation of pinning
energies corresponding to the parameters of the inner crust
(Table~1) have been performed in the framework of the Local
Density Approximation (LDA); details of the model are given in
Ref.~\refcite{DP1}.
\begin{figure}[b]
\begin{center}
  \parbox{2.4in}{\epsfig{figure=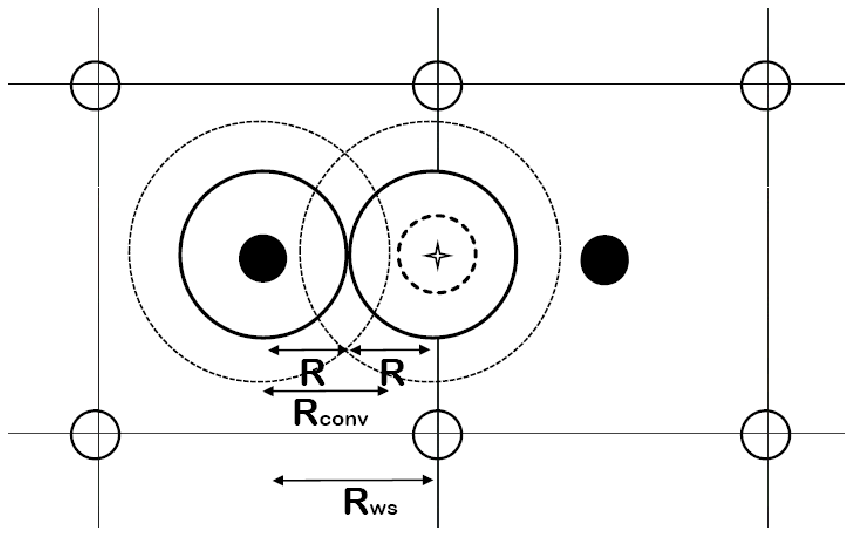,width=2.4in}}
  \hspace*{6pt}
  \parbox{2.4in}{\epsfig{figure=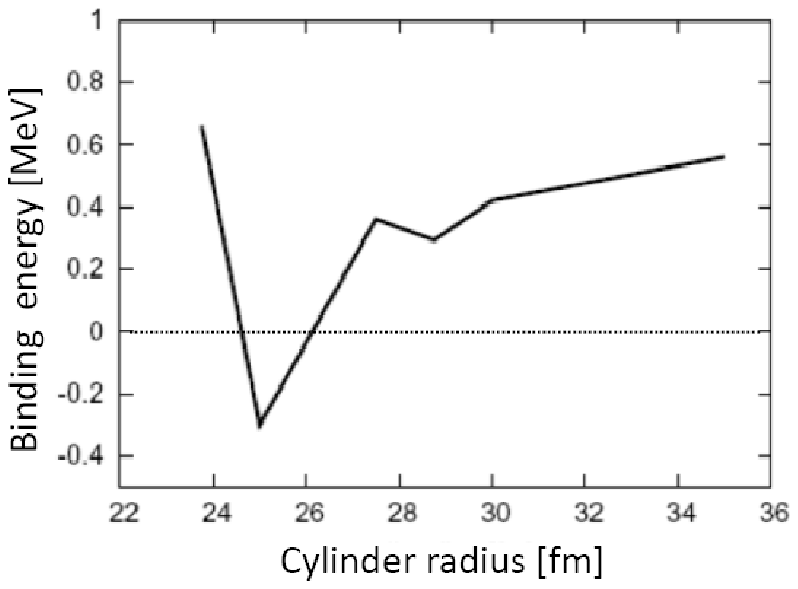,width=2.4in}}
  \caption{Left: lattice in the dense limit $R_{_{\rm WS}} < 2 R_{_{\rm
conv}} $.  \  Right: dependence of the
  binding energy on the radius of the quantization cylinder (figure taken from
  Ref.~\refcite{orsay}). }
\end{center}
\end{figure}
The purely additive nature of energy contributions from adjacent
volumes of matter in this semi-classical approximation allows on
the one hand to deal with the difficult geometry of pinning; but
on the other hand,  it completely neglects 'proximity' effects
associated to the non-local quantum nature of pairing phenomena.
In the absence of credible quantum calculations for the
vortex-lattice system, however, which can confirm or modify the
semi-classical predictions,  the LDA results are the only
physically reasonable input for the vortex theory of glitches.

Binding energies are relevant to glitch theory in two related
aspects:  they allow to determine the convergence distance
$D_{_{\rm conv}} = 2 R_{_{\rm conv}} $  and they are equal to the
pinning energies for the density regions where the lattice is
dilute, i.e. where $R_{_{\rm WS}} > D_{_{\rm conv}}$. Recently, a
quantum calculation of binding energies has been performed by
solving the mean-field Hartree-Fock-Bogoliubov (HFB) equations.
Although improperly and quite misleadingly called pinning energy,
the quantity calculated in Ref.~\refcite{ABBV} is obtained
precisely as shown in Figure~2~(right), and convergence is
correctly required as a consistency condition for the model. In
Figure~4~(right) we show their study of the convergence of $E_{\rm
bind}$ as a function of the cylinder radius $R$. At the value
$R=30$ fm used in Ref.~\refcite{ABBV}, \emph{reasonable}
convergence is finally reached (the still increasing  $E_{\rm
bind}$ is included in the error bars in their figures). Being
optimistic one may go down to $R_{_{\rm conv}}\approx 27-28$ fm,
but lower values  present wild oscillations (at $R\sim 25$ fm,
$E_{\rm bind}$ even reverses its sign), obviously out of control
and not related to any physics but only to the artificial boxes of
the quantization procedure (the 'dripped' neutrons occupy
positive-energy \emph{continuum} states). One thus finds $D_{_{\rm
conv}}\approx 55$ fm, and thence the dilute limit corresponds to
$R_{_{\rm WS}} \gtrsim 55 $ fm. The results of Ref.~\refcite{NV}
show that this condition corresponds to only the outermost layers
of the inner crust, with densities $\rho_{_{\rm B}} \lesssim 4.7
\times 10^{11}$ g/cm$^{3}$ and $n_{_{\rm G}} \lesssim 10^{-4}$
fm$^{-3}$ ($k_{_{\rm F,G}} \lesssim 0.14$ fm$^{-1}$). This is very
unwelcome, since it does not allow to use binding in place of
pinning in most of the crust, particularly in the density regions
which appears to be relevant to glitches in the LDA, i.e. around
zone~3 \cite{DP1}.

 In Figure~5~(left) we report the comparison between LDA and HFB
 results used  in Ref.~\refcite{ABBV}  to support their main
conclusion, namely that around zones~3 and 4 pinning is not
nuclear, but interstitial. This conclusion is wrong and
misleading: as shown by the vertical line at $k_{_{\rm F,G}} =
0.14$ fm$^{-1}$, the figure compares two different energies in a
regime where they do not represent the same quantity. To state it
more vividly, in order to describe pinning in  zone~1 ($R_{_{\rm
WS}} = 44$ fm)  the HFB approach should use cylinders with $R=22$
fm and the result should already converge at this small
 radius; to describe zone~3, convergence should be obtained
at $R=14$ fm!

Where the dilute limit is reached, however, the HFB results for
$E_{\rm bind}$ can and should be compared to the LDA results for
$E_{\rm pin}$, in order to assess whether the semi-classical model
is realistic or not, at least in this regime. One can actually
exploit the 'locality' of the LDA to extend the analysis a bit
further. Indeed, the absence of proximity effects is such that,
when applied to the calculation of $E_{\rm bind}$, the LDA
converges to a definite final value already for $D_{_{\rm
conv}}\approx 30$ fm. Therefore, at densities $n_{_{\rm G}}
\lesssim 0.01 $ fm$^{-3}$ ($k_{_{\rm F,G}} \lesssim 0.6$
fm$^{-1}$), where $R_{_{\rm WS}}\gtrsim 30$ fm, the LDA values for
pinning are the same as its results for binding and as such they
could be compared to the corresponding HFB results. However,
before claiming that at very low densities (between zones~1 and 2)
the HFB and LDA predictions for binding have opposite sign, like
Figure~5~(left) may suggest, some crucial issues should be kept in
mind: \emph{(i)} Figure~5~(right) shows  the HFB results for
different pairing interactions; the corresponding binding energies
between zones~1 and 2 can differ even in the sign. Thence the
comparison must be done with the 'same' pairing interaction, in
the sense that the  density dependence of the neutron gap in
uniform matter must be the same in HFB and LDA. Instead, the
interaction used in Ref.~\refcite{ABBV} is equivalent to Gogny
D1S, which for $k_{_{\rm F,G}} > 1$ fm$^{-1}$ yields larger gaps
than the Argonne used in the LDA.  \emph{(ii)} As carefully
explained in Ref.~\refcite{DP1}, the pinning energy should be
evaluated at \emph{fixed chemical potential} (the particle bath
represented by the macroscopic inner crust which surrounds the
widely spaced vortices), while the results of Ref.~\refcite{ABBV}
correspond to a fixed number of particles. Since $E_{\rm bind}$
comes from the difference between very large numbers, great
attention must be devoted to this kind of issue. \emph{(iii)}
Different zones present nuclei with different radii, while in
Ref.~\refcite{ABBV} the same nucleus (with $R_{_{\rm N}} = 7 $ fm)
was taken at all densities.

With these issues under control, the comparison of quantum and
semi-classical results for binding at low densities could become
quite instructive.

\begin{figure}[t]
\begin{center}
  \parbox{2.4in}{\epsfig{figure=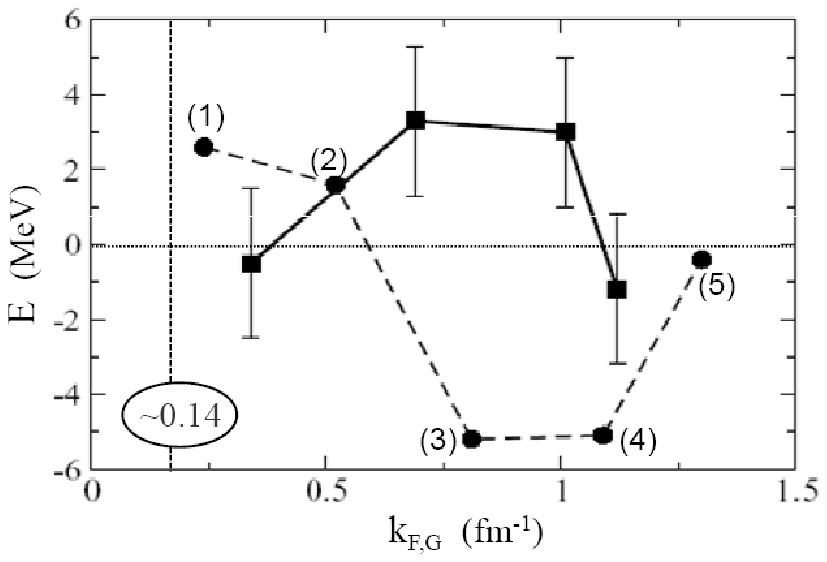,width=2.4in}}
  \hspace*{6pt}
  \parbox{2.4in}{\epsfig{figure=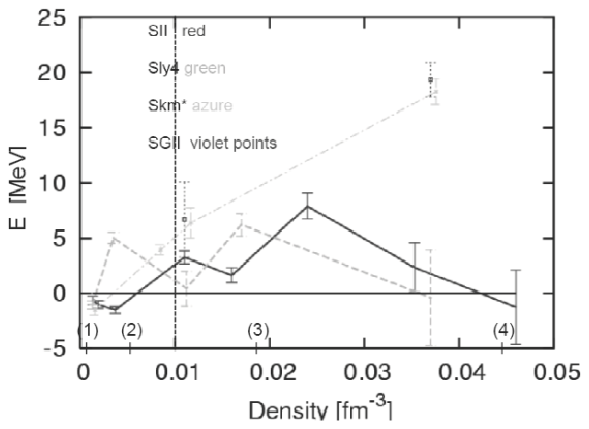,width=2.4in}}
  \caption{Left: comparison of LDA results for pinning energies (dashed line) and HFB results
  for binding energies (solid line); the numbers indicate the zones of Table~1 (figure taken from Ref.~\refcite{ABBV}). \ Right: comparison of HFB
  results for binding energies corresponding to different choices of the
  pairing interaction (figure taken from   Ref.~\refcite{orsay}). }
\end{center}
\end{figure}

\end{document}